\documentclass[sn-mathphys-num]{sn-jnl}

\usepackage{graphicx}%
\usepackage{multirow}%
\usepackage{amsmath,amssymb,amsfonts}%
\usepackage{amsthm}%
\usepackage{mathrsfs}%
\usepackage[title]{appendix}%
\usepackage{xcolor}%
\usepackage{textcomp}%
\usepackage{manyfoot}%
\usepackage{booktabs}%
\usepackage{algorithm}%
\usepackage{algorithmicx}%
\usepackage{algpseudocode}%
\usepackage{listings}%
\usepackage{diagbox,multirow,subcaption,tabularx,multicol}
\usepackage[detect-all=true,group-minimum-digits=4]{siunitx}
\usepackage{colortbl,array}
\usepackage{wrapfig}
\usepackage{longtable,hhline,ifthen,url,booktabs}

\usepackage[version=4]{mhchem}

\usepackage{cleveref}

\newcolumntype{Y}{>{\small\raggedright\arraybackslash}X}

\newcommand{\mol}{\textbf{\ce{[NH3-(CH2)4-NH3]CdCl4}}~}
\newcommand{\ionN}{\textbf{\ce{[CdCl4]^{2-}}}~}
\newcommand{\ionP}{\textbf{\ce{[NH3-(CH2)4-NH3]^{2+}}}~}
\newcommand{\PhI}{\textbf{Phase I}~}
\newcommand{\PhV}{\textbf{Phase V}~}

\DeclareSIUnit\angstrom{\text{Å}} 

\begin{document}

\title[First principal investigation of Structural optical and thermoelectric
properties of hybrid organic-inorganic perovskite \ce{[NH3-(CH2)4-NH3]CdCl4} compound]{First principal investigation of Structural optical and thermoelectric
properties of hybrid organic-inorganic perovskite \ce{[NH3-(CH2)4-NH3]CdCl4} compound}


\author[1]{\fnm{Hafida} \sur{ZIOUANI}}\email{ha.ziouani@edu.umi.ac.ma}

\author[1]{\fnm{Sanaa} \sur{MAZOUAR}}\email{s.mazouar@edu.umi.ma}

\author*[2]{\fnm{Jean-Pierre} \sur{TCHAPET NJAFA}}\email{jean-pierre.tchapet@facsciences-uy1.cm}

\author[1]{\fnm{Taoufik} \sur{ABDELILAH}}\email{taoufik.abdelillah@edu.umi.ac.ma}

\author[1]{\fnm{Mahmoud} \sur{ETTAKNI}}\email{M.ETTAKNI@umi.ac.ma}

\author[1]{\fnm{El Mostafa} \sur{KHECHOUBI}}\email{khechoubi@umi.ac.ma}

\affil[1]{\orgdiv{Department of Physics}, \orgname{Faculty of Science, Materials and Renewable Energy Team, LP2MS Laboratory, Moulay Ismail University}, \orgaddress{\city{Meknes}, \country{Morocco}}}

\affil*[2]{\orgdiv{Department of Physics}, \orgname{Faculty of Science, University of Yaoundé 1}, \orgaddress{\city{Yaoundé}, \postcode{812}, \country{Cameroon}}}

\abstract{The structural, thermoelectric, and optical properties of \mol
were studied using Density Functional Theory (DFT) within the ABINIT code. The
GGA-PBE functional, plane wave pseudopotentials, a kinetic energy cutoff of $\qty{35}{Ha}$,
and an 11x8x8 Monkhorst-Pack $k$-point grid were employed. The material comprises
inorganic \ionN sheets, organic \ionP layers, and \ce{N-H-Cl}
hydrogen bonds, ensuring sublattice cohesion. Structural optimization used reference
crystal data, enabling analysis of alkylene-diammonium chain conformation,
intermolecular interactions, and crystal stability. The study highlights the role of Cd in
influencing optical and thermoelectric properties. Temperature-induced changes lead
to a reduced band gap and enhanced optical absorption, indicating significant
electronic structure modifications. These findings propose \mol
as a promising candidate for optoelectronic applications, particularly after thermal
cycling, due to its improved performance under varying conditions.}





\maketitle

\section{Introduction}
Due to their numerous applications in technology and industry, perovskite compounds
have been the subject of considerable interest among researchers. The structure of perovskite
materials was first discovered in 1839 by Gustav Rose, and these materials exhibit a range of
distinctive and exceptional physical properties, including ferromagnetism, spin-dependent
transport, high $Tc$ superconductivity, high thermopower, ferroelectricity, and so on.

Given their abundance in nature, low cost, and ease of deposition into thin films, organic-
inorganic halide hybrid perovskites (\ce{ABX3}: \ce{A} = big organic cation, \ce{B} = metal cation, and \ce{X} =
halogen anion) offer considerable potential for usage in commercial applications \cite{Borriello2008}.

Thermoelectric materials are of universal interest due to their capacity to transform heat
into electric energy, rendering them a significant source of energy \cite{zhang2015,Metal2021}. Currently, the world
requires thermoelectric supplies to reduce its reliance on fossil fuels. Perovskite compounds
exhibit exceptional thermoelectric properties, which have prompted the development of
thermoelectric device appliances. In the field of high-temperature thermoelectric research,
germanium-based chloro perovskites have garnered significant attention in recent years \cite{nyayban2021}.
Lead is a toxic substance for the environment, whereas hybrid perovskite is a free material and
therefore environmentally benign. Consequently, it can be employed in photovoltaic
applications with minimal environmental impact.

The crystal structure of \mol is composed of an infinite number of
inorganic layers. Each layer is formed by corner-sharing \textbf{\ce{CdCl6}} octahedra, arranged in a two-
dimensional perovskite-like structure. The alkyl ammonium (or alkylenedi ammonium) chains
are oriented almost perpendicular to the planes of the layer. The \ce{NH3} groups at the end of the
organic chain engage in \textbf{\ce{N-H-Cl}} hydrogen bonds with the octahedra.

Previous studies have investigated thermal and crystallographic analyses \cite{KHECHOUBI1994}. Evidence
of five crystalline forms, designated V, IV, III, II, and I, has been identified. It has been
established that the various phase transitions are either irreversible or reversible. The various
phase transitions of this complex can be associated with the conformational evolutions of the
organic part
\ionP, thanks to structural determinations of the "limit" Phases
five (\textbf{V}: the initial room-temperature form) and one (\textbf{I}: the new room-temperature form) \cite{Reiss2013}. A
review of the literature reveals that an increase in temperature at \PhV results in an
irreversible change from the "gauche" conformation of the organic chain to an "all-trans"
conformation, with the final conformation retained during subsequent phase transitions. In addition to these crystallographic studies, recent work has 
investigated the electronic and optical properties of similar hybrid perovskites, such as \ce{[NH3-(CH2)4-NH3]MX4} (\ce{M} = \ce{Cu}, \ce{Mn}), using Density 
Functional Theory (DFT) \cite{Hafida2024}. These studies have revealed that the substitution of \ce{Cu} and \ce{Mn} significantly alters the electronic and 
optical behaviors, making these compounds promising candidates for optoelectronic applications. 

In this work, we will study the effect of Phase transitions, especially the \PhV and \textbf{I},
on the structural, electronic, optical, and thermoelectric properties of the compound \mol.

\section{Computational details}

The Density Functional Theory+U (DFT+U) method implemented in the ABINIT code \cite{Gonze2002,Gonze2005}
was utilized to investigate the structural and electronic properties of \mol.
The double counting correction known as Full localised limit (FLL) \cite{Liechtenstein1995} was employed, along
with the Generalised Gradient Approximation (GGA) in the Perdew-Burke-Ernzerhof
functional \cite{Perdew1997}. The DFT+U method is an extension of the standard Density Functional
Theory (DFT) approach that incorporates an additional correction for treating the electron-
electron interactions within localized orbitals. The pseudopotential was generated using the
Projector Augmented-Wave approach \cite{Holzwarth2001}. A cutoff of kinetic energy at $\qty{35}{Ha}$ was utilized to
carry out self-consistent field (SCF) and non-self-consistent field (NSCF) calculations on \mol. The Monkhorst Pack Mesh scheme adopted a $k$-points grid 
sampling of
11x8x8 for conducting the integrations of the irreducible Brillouin zone of \mol. The crystal data of \mol, as reported in the literature \cite{KHECHOUBI1994},
served as the initial reference point.

\section{Results and discussion}
\subsection{Structural properties}

\begin{figure}[htbp]
 \centering
 \leavevmode
 \subfloat[\PhV]{\includegraphics[width=0.5\textwidth]{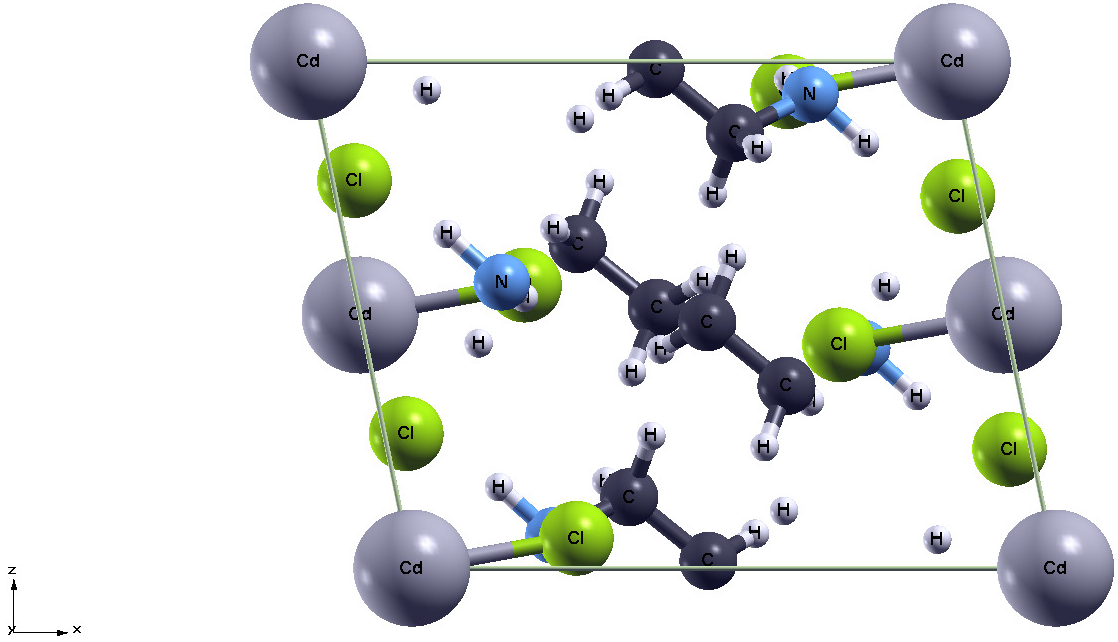}}\hfill
 \subfloat[\PhI]{\includegraphics[width=0.5\textwidth]{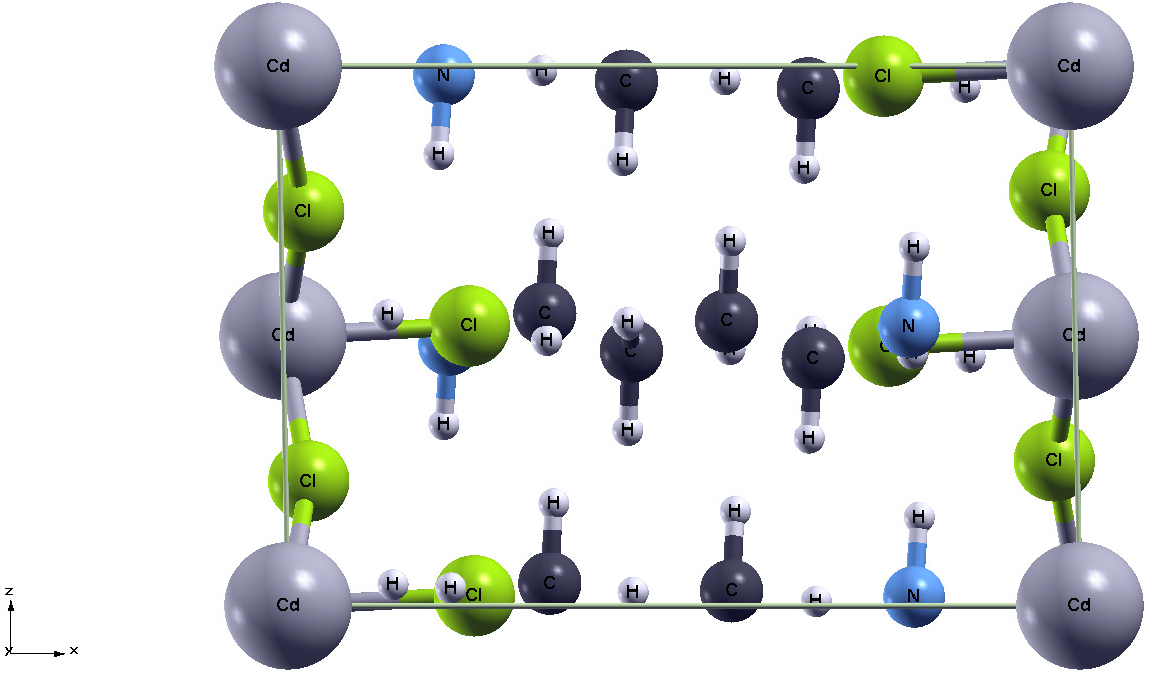}}
 \caption{Crystal structure of \mol.}\label{fig-struc}
\end{figure}

After a geometry optimization, the fitting of Birch-Murnaghan equation of state \cite{houmad2015,Murnaghan1944}
results in optimized lattice parameters of \mol by the GGA-PBE
approximation in the two Phases are presented in \Cref{fig-struc,tab-compound}, which are in great agreement with
previous experimental measurements findings \cite{KHECHOUBI1994}, confirming the accuracy of the present
simulation. More interestingly, all the properties treated in this work (electronic, optical, and
thermoelectric properties) are calculated after the full structure relaxation, including lattice
constants, atomic positions, and free total energies of supercells.

\begin{table}[htbp]
\caption{Parameter of the two Phases of \mol.}\label{tab-compound}
{%
\newcommand{\mc}[3]{\multicolumn{#1}{#2}{#3}}
\begin{tabularx}{0.9\textwidth}{m{3cm}YY}
\toprule
\multirow{3}{*}{\diagbox[innerwidth=3cm,height=3\line]{Crystal-\\line system}{Crystal lattice\\structure}} & \mc{1}{c}{Monoclinic}& \mc{1}{c}{Monoclinic}\\
 & \PhV: initial room-temperature form & \PhI: new room-temperature form\\\midrule
\mc{1}{c}{$a$} & \mc{1}{c}{$\qty{7.657}{\angstrom}$} & \mc{1}{c}{$\qty{7.344}{\angstrom}$}\\
\mc{1}{c}{$b$} & \mc{1}{c}{$\qty{7.585}{\angstrom}$} & \mc{1}{c}{$\qty{7.485}{\angstrom}$}\\
\mc{1}{c}{$c$} & \mc{1}{c}{$\qty{9.541}{\angstrom}$} & \mc{1}{c}{$\qty{10.775}{\angstrom}$}\\
\mc{1}{c}{$\beta$} & \mc{1}{c}{$\qty{101.560}{\degree}$} & \mc{1}{c}{$\qty{91.060}{\degree}$} \\
\mc{1}{c}{$Z$} & \mc{1}{c}{\num{2}} & \mc{1}{c}{\num{2}}\\
\mc{1}{c}{Space group} & \mc{1}{c}{P2$_1/a$} & \mc{1}{c}{P2$_1/a$}\\\bottomrule
\end{tabularx}
}%
\end{table}

\subsection{Electronic properties}

\begin{figure}[htbp]
\centering
\leavevmode
\subfloat[\PhV]{\includegraphics[width=0.5\textwidth]{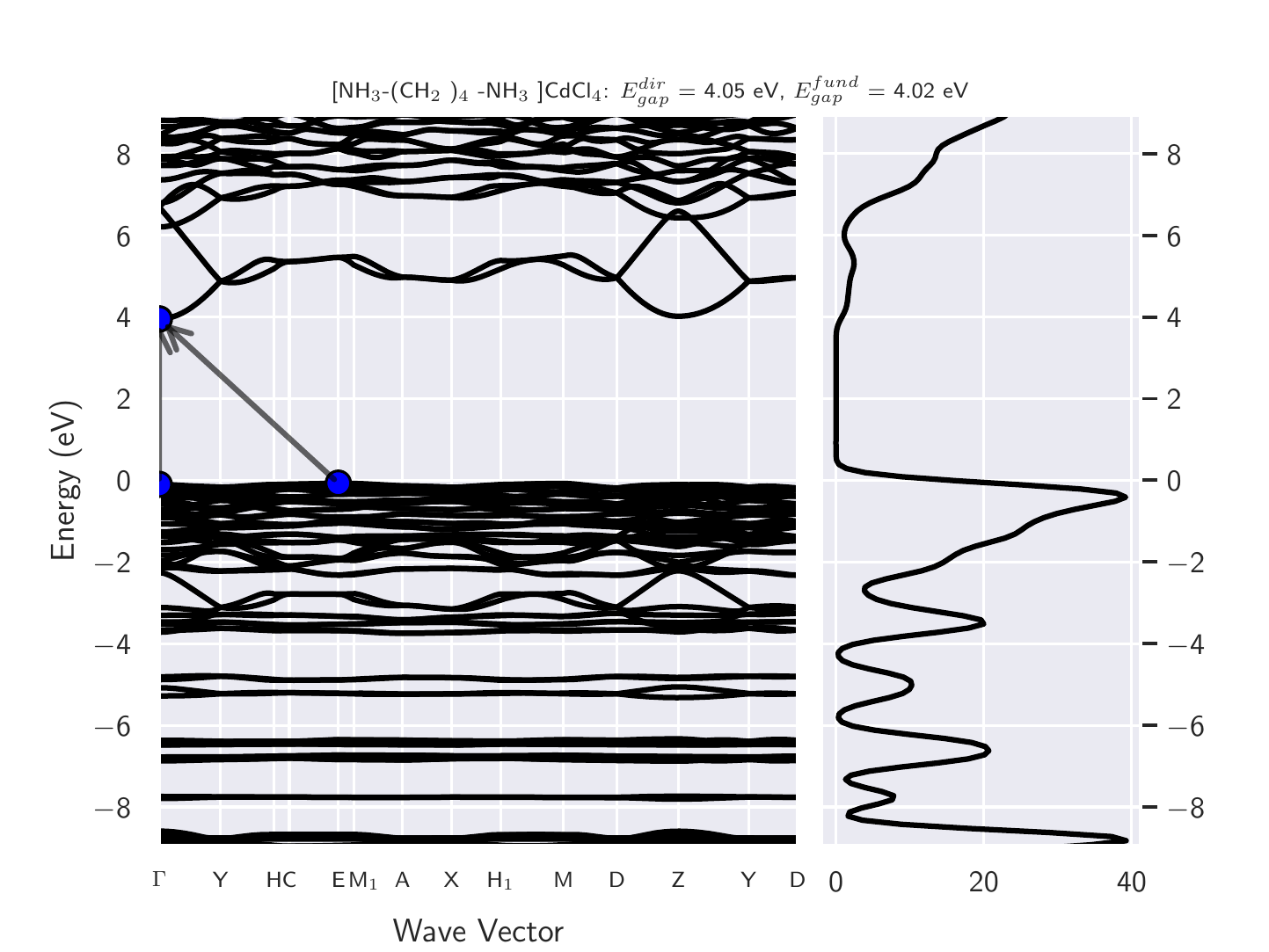}}\hfill
\subfloat[\PhI]{\includegraphics[width=0.5\textwidth]{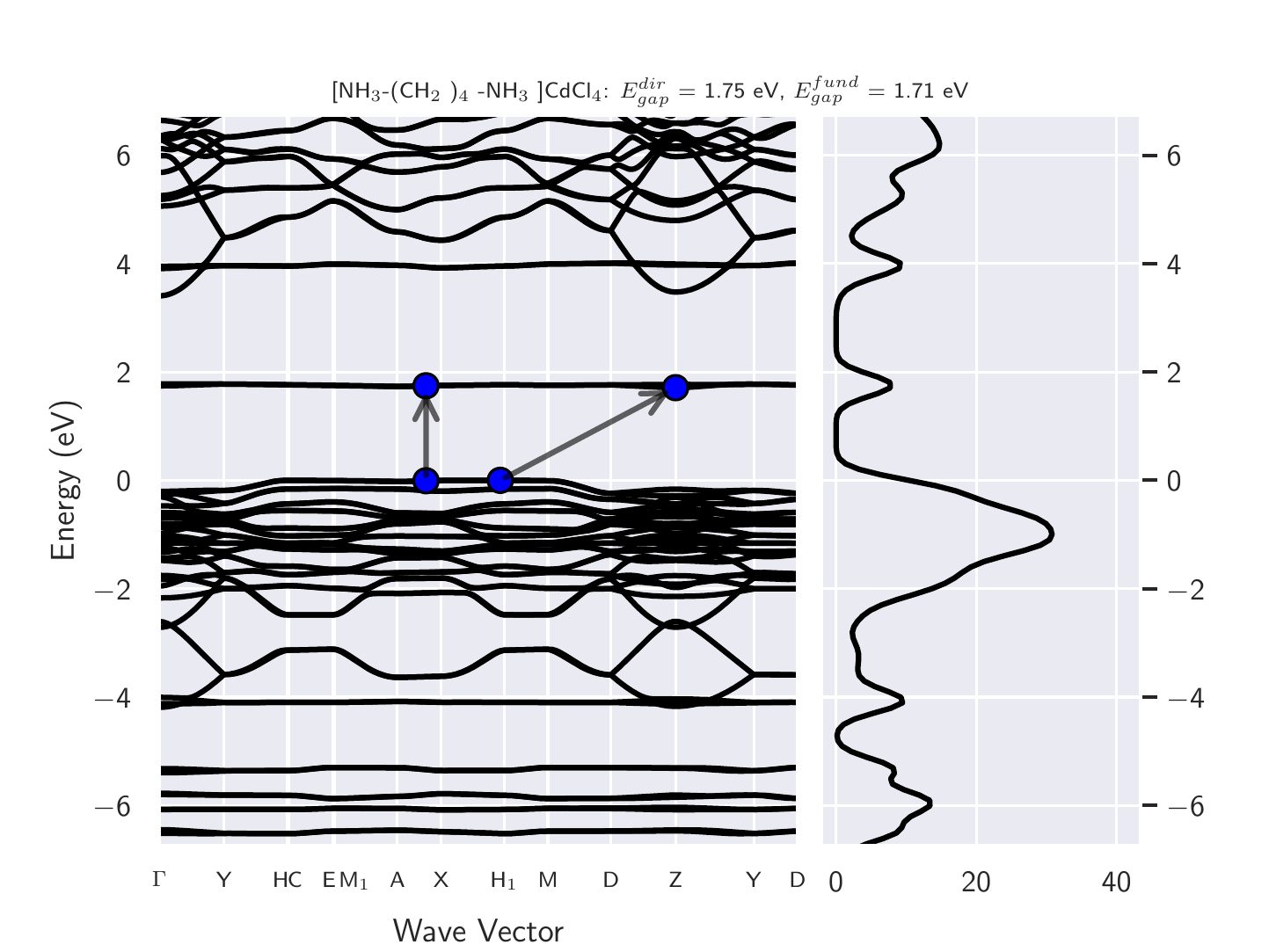}}
\caption{Electronic band structures of \mol.}\label{fig-bandStruc}
\end{figure}

The density of the state is one of the most important properties that gives information
about the behavior and electronic character of this system. It also allows knowing the nature of
the chemical bonds between atoms. The total (TDOS) and partial (PDOS) densities of states are
calculated at their equilibrium states by the GGA-PBE approach. The projected results between
$\qty{-12}{\electronvolt}$ and $\qty{12}{\electronvolt}$ are shown in \Cref{fig-bandStruc}. The Fermi level is taken at the energy of $\qty{0}{\electronvolt}$.

To obtain the precise band gap (see \Cref{tab-ener}) of this compound, the generalized gradient
approximation (GGA-PBE) was used.

\begin{table}[htbp]
\caption{Gap energy values obtained by DFT+U method.}\label{tab-ener}
\begin{tabular}{cccc}
\toprule
Crystal structure & $E_g(\unit{\electronvolt})$ & Fermi level $(\unit{\electronvolt})$ & Gap type\\\midrule
\PhV & \num{4.020} & \num{-0.463} & indirect\\
\PhI & \num{1.710} & \num{0.074} & indirect\\\bottomrule
\end{tabular}
\end{table}

\begin{figure}[htbp]
\centering
\leavevmode
\subfloat[\PhV]{\includegraphics[width=0.5\textwidth]{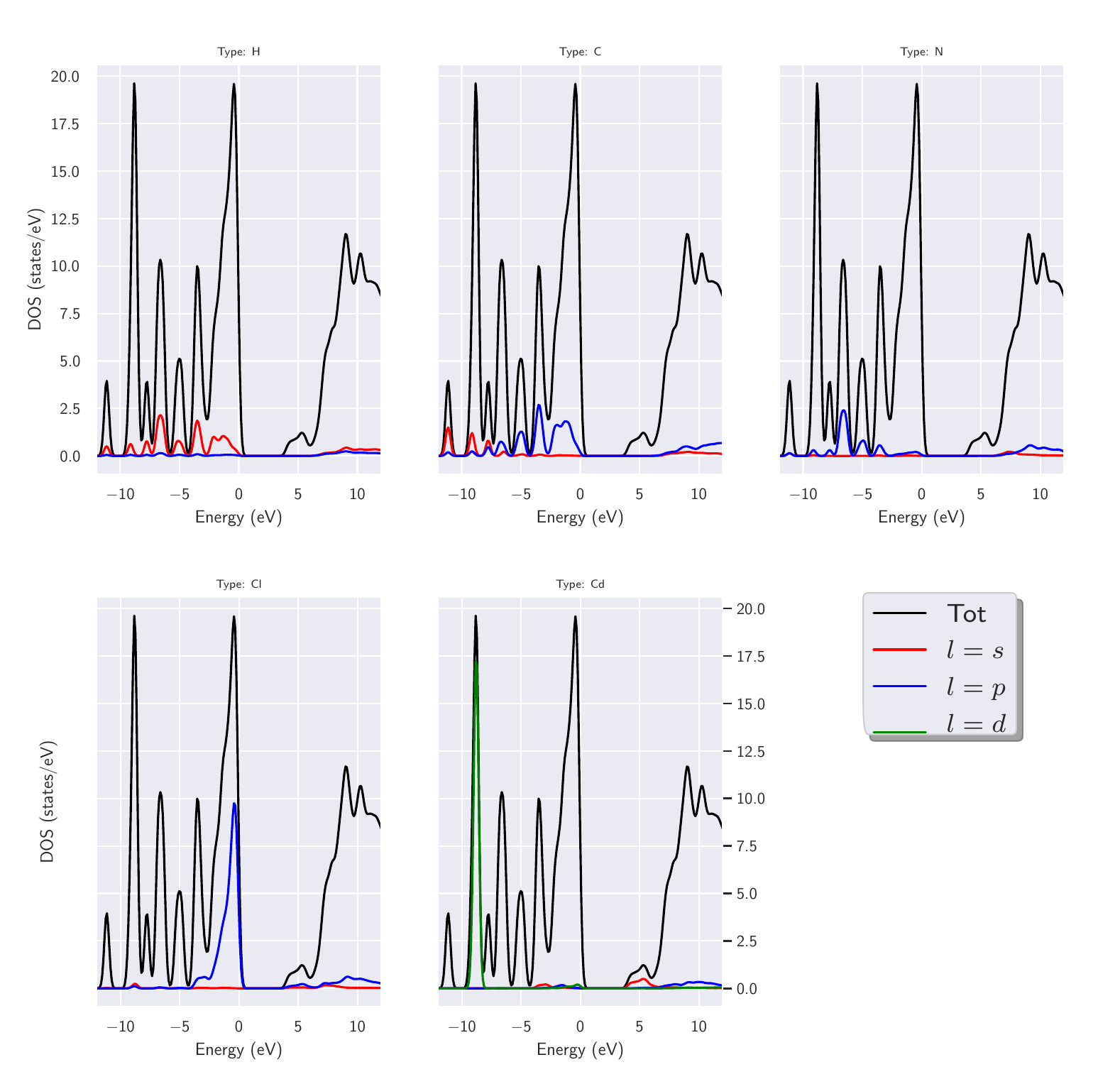}}\hfill
\subfloat[\PhI]{\includegraphics[width=0.5\textwidth]{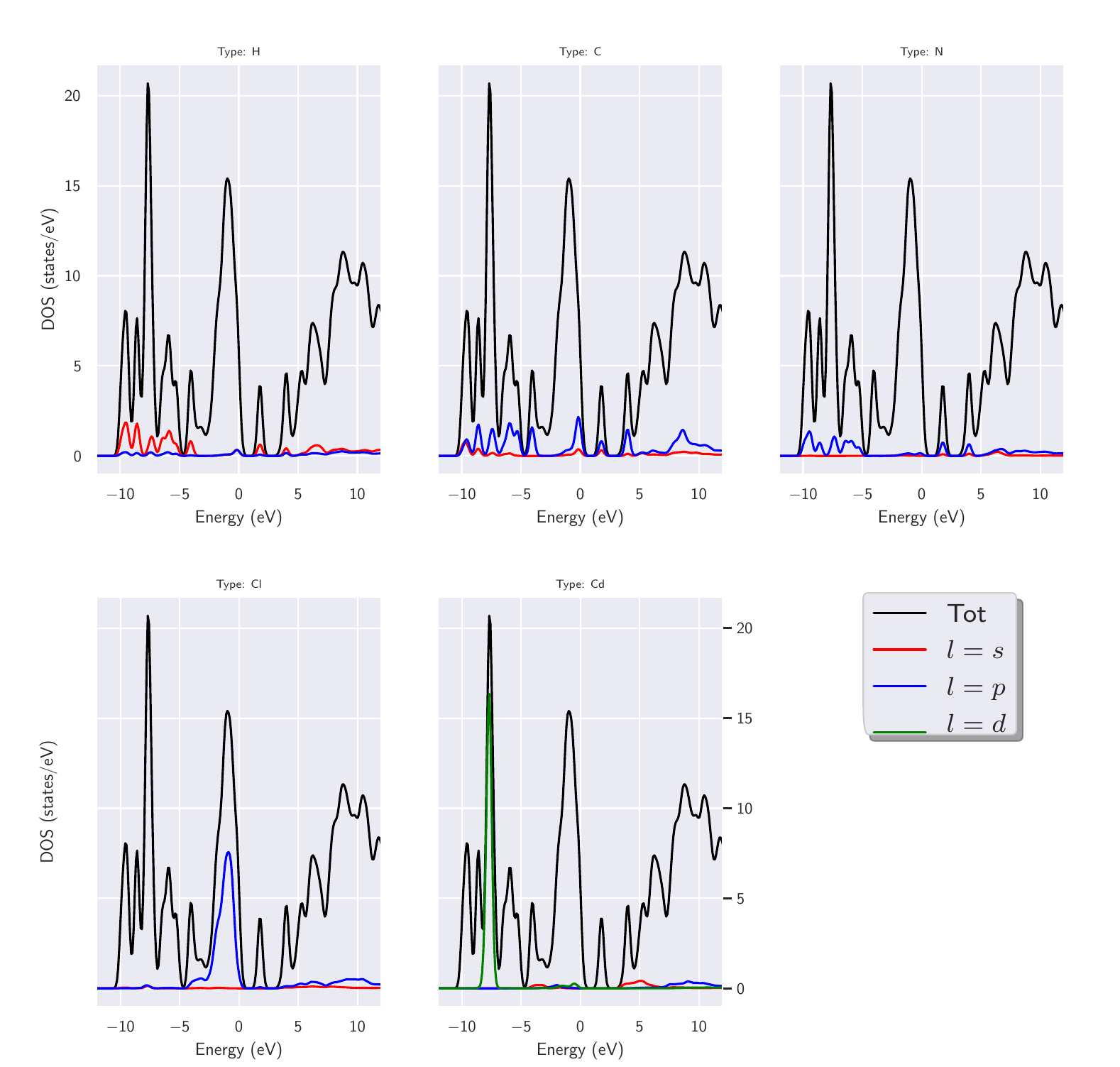}}
\caption{Total and partial density of states of \mol.}\label{fig-DensSt}
\end{figure}

\subsubsection{Case of the \PhV}

Analysis of the density of states (DOS) plot (see \Cref{fig-bandStruc,fig-DensSt}) pertaining to \PhV of \mol offers valuable insights into 
the orbital contributions of each
constituent element (\ce{H}, \ce{C}, \ce{N}, \ce{Cl}, \ce{Cd}) to the overall electronic configuration.

\begin{itemize}
 \item Hydrogen (\ce{H}): Examination of the DOS plot implies that the $s$-orbital plays a
predominant role across the complete energy spectrum displayed, which is consistent
with hydrogen's lone valence electron primarily occupying the $s$-orbital.
\item  Carbon (\ce{C}): The primary contributor for carbon appears to be the $p$-orbital, exhibiting
peaks throughout the spectrum, while the contribution from the $s$-orbital, if any, seems
negligible compared to the $p$-orbitals. This observation indicates that carbon's four
valence electrons predominantly reside in its $p$-orbitals.
\item Nitrogen (\ce{N}): Analogous to carbon, the dominant contributor for nitrogen seems to be
the p-orbital, evident through peaks spanning the entire spectrum. The presence of the
$s$-orbital contribution, if existent, appears minor considering the scale, in accordance
with nitrogen's five valence electrons predominantly occupying its $p$-orbitals.
\item Chlorine (\ce{Cl}): Once again, the major contribution for chlorine is attributed to the
$p$-orbital, with peaks distributed across the spectrum. The potential presence of the 
$s$-orbital contribution seems minor, reflecting chlorine's seven valence electrons primarily
filling its $p$-orbitals as indicated in the DOS plot.
\item Cadmium (\ce{Cd}): A significant contribution throughout the spectrum is observed from
the $d$-orbital for cadmium, with the possibility of a presence of the $p$-orbital
contribution, albeit weaker than the $d$-orbitals. The contribution from the $s$-orbital is
anticipated to be minimal based on the scale, aligning with cadmium's ten valence
electrons occupying its $4d$-orbitals and partially populating the $5p$-orbitals.
\end{itemize}

The identification of a band gap (refer to \Cref{fig-DensSt})
($E^{dir}_g=\qty{4.050}{\electronvolt}$, $E^{fund}_g=\qty{4.200}{\electronvolt}$) signifies the
semiconductor nature of \PhV of \mol. The wide band gap observed
suggests potential utilization in optoelectronic devices designed for operation in the ultraviolet
(UV) region.

In summary, the analysis of the DOS for the \PhV of \mol indicates a
predominant role of $p$-orbitals for \ce{H}, \ce{N}, and \ce{Cl}, with a substantial contribution from $p$-orbitals
for carbon as well. Cadmium, characterized by filled d-orbitals, exhibits a significant
contribution from $d$-orbitals to the DOS. The evaluation of the band gap validates the
semiconductor characteristics of this phase, demonstrating a broad band gap and positioning it
as a promising contender for UV optoelectronic applications.

\subsubsection{Case of the \PhI}

Like before (see \Cref{fig-bandStruc,fig-DensSt}), the analysis of the partial density of states (PDOS) plot for
the \PhI of \mol yields valuable insights into the orbital contributions
of each constituent element to the total density of states (DOS).

\begin{itemize}
 \item Hydrogen (\ce{H}): Examination of the PDOS plot affirms the dominance of the $s$-orbital
below the Fermi level with minimal contributions above it, consistent with the fact that
hydrogen's single valence electron primarily occupies the s-orbital. The marginal $p$-
orbital contribution corresponds to expectations.
\item Carbon (\ce{C}): The $p$-orbital demonstrates significant contributions with notable peaks on
either side of the Fermi level, while the $s$-orbital contribution is present but relatively
smaller. This observation implies that carbon's four valence electrons predominantly
occupy its $p$-orbitals, although some involvement of s-orbitals is also evident.
\item Nitrogen (\ce{N}): Analogous to carbon, nitrogen exhibits a dominance of $p$-orbital
contribution, with prominent peaks both below and above the Fermi level. The $s$-orbital
contribution, although present, is comparatively minor, consistent with nitrogen's five
valence electrons primarily occupying its $p$-orbitals.
\item Chlorine (\ce{Cl}): Once again, the $p$-orbital manifests a significant contribution with peaks
on both sides of the Fermi level, while the $s$-orbital contribution is minor. Chlorine's
seven valence electrons primarily occupy its $p$-orbitals, as indicated in the PDOS plot.
\item Cadmium (\ce{Cd}): Notably, the d-orbital displays a substantial contribution with
significant peaks below and above the Fermi level. The p-orbital contribution is
moderate, whereas the $s$-orbital contribution is minor, aligning with the presence of ten
valence electrons in cadmium, occupying its $4d$-orbitals and partially the $5p$-orbitals.
\end{itemize}

The computed direct band gap ($E^{dir}_g=\qty{1.750}{\electronvolt}$) and fundamental band gap ($E^{fund}_g = \qty{1.710}{\electronvolt}$)
deduced from the band structure plot indicate the semiconducting nature of the \PhI of
\mol. The presence of bands intersecting the Fermi level further
substantiates the existence of electronic states in proximity to the band gap, potentially
impacting electrical conductivity and other material properties.

In short, the examination of PDOS for the \PhI of \mol elucidates the
pivotal role of $p$-orbitals in determining the DOS for \ce{H}, \ce{N}, and \ce{Cl}. Carbon also exhibits a
noteworthy contribution from $p$-orbitals, accompanied by some involvement of $s$-orbitals.
Cadmium, characterized by filled $d$-orbitals, showcases a substantial contribution from 
$d$-orbitals to the DOS. The band gap analysis corroborates the semiconductor properties of the
material. This profound comprehension of the electronic structure can offer valuable insights
for customizing material properties for optoelectronic applications.

\subsection{Optical properties}

The study of the optical properties of solids has proven to be a powerful tool in
understanding the electronic properties of materials. In this section, the study of optical
properties is carried out using the GGA-PBE approximation, which has been shown to be
successful in determining the optics band gap with appreciable precision. The optical properties
of any material can be described by the complex dielectric function (see \Cref{eq-die}) \cite{houmad2015}:

\begin{equation}
\varepsilon(\omega)=\varepsilon_1(\omega)+i\varepsilon_2(\omega).
 \label{eq-die}
\end{equation}

It describes the optical response of the medium at all photon energies $E=\hbar\omega$. The real part of the dielectric function follows from the 
Kramers–Kronig relation \eqref{eq-KK} \cite{houmad2015}:
\begin{equation}
\varepsilon_1(\omega)=1+\frac{2}{\pi}P\int\limits_0 ^\infty\frac{\varepsilon_2(\alpha)}{\alpha^2-\omega^2}\alpha d\alpha,
 \label{eq-KK}
\end{equation}
where $P$ implies the principal value of the integral.

The imaginary part, $\varepsilon_2(\omega)$, in the long wavelength limit, has been obtained directly from the electronic structural calculation (see 
\Cref{eq-es}) \cite{KHENATA2006}:
\begin{equation}
\varepsilon_2(\omega)=\frac{e^2\hbar}{\pi m^{2\omega^2}}\sum_{v, c}\int\limits_{BZ}\left|M_{cv}(k)\right|^2\delta\left[\omega_{cv}(k)-\omega\right]d^3k,
 \label{eq-es}
\end{equation}
with $M_{cv}(k)$ the transition moment elements. 

The following formula \eqref{eq-ac} is employed to calculate the absorption coefficient \cite{houmad2015,KHENATA2006,HOUMAD2016}:
\begin{equation}
 \alpha(\omega)=\frac{\sqrt{2}}{c}\omega\sqrt{-\varepsilon_1(\omega)+\sqrt{\left(\varepsilon_1^2(\omega)+\varepsilon_2^2(\omega)\right)}}.
 \label{eq-ac}
\end{equation}
The optical conductivity is given by the \Cref{eq-oc} \cite{Perdew1997}:
\begin{equation}
 \mathtt{Re}\left(\sigma_{\alpha\beta}(\omega)\right)=\frac{\omega}{4\pi}\mathtt{Im}\left(\varepsilon_{\alpha\beta}(\omega)\right).
 \label{eq-oc}
\end{equation}

The calculation of reflectivity is very important to evaluate the optical properties of materials; it makes it easy to understand the status and behavior of 
transparency, based on the reflectivity that we can draw from the transmittance. The reflectivity $R(\omega)$ (see 
\Cref{eq-re}) follows directly from the Fresnel’s formula \cite{KHENATA2006}:
\begin{equation}
R(\omega)=\left|\frac{\sqrt{\varepsilon(\omega)}-1}{\sqrt{\varepsilon(\omega)}+1}\right|^2.
 \label{eq-re}
\end{equation}

It is also recommended to calculate the energy versus the wavelengths by varying the
refractive index. This allows for an understanding of how the material can behave and how it
can be canalized. To confirm the results found by the optical absorption and the transmittance,
the refractive index $n(\omega)$ is given as follows in \Cref{eq-ri} \cite{houmad2015}:
\begin{equation}
 n(\omega)=\sqrt{\frac{\sqrt{\varepsilon_1^2(\omega)+\varepsilon_2^2(\omega)+\varepsilon_1(\omega)}}{2}}.
 \label{eq-ri}
\end{equation}

This section presents a study of the effect of temperature on the optical properties of \mol, including reflectivity and optical conductivity. Due to symmetry, 
the
optical properties differ for the three directions, $xx$, $yy$ and $zz$. The subsequent analysis will
focus on the $xx$ direction which is the most favorable for light production (the same
consideration is also made for subsection \ref{subsec-Ther}).

\begin{figure}[htbp]
\centering
\leavevmode
\subfloat{\includegraphics[width=0.5\textwidth]{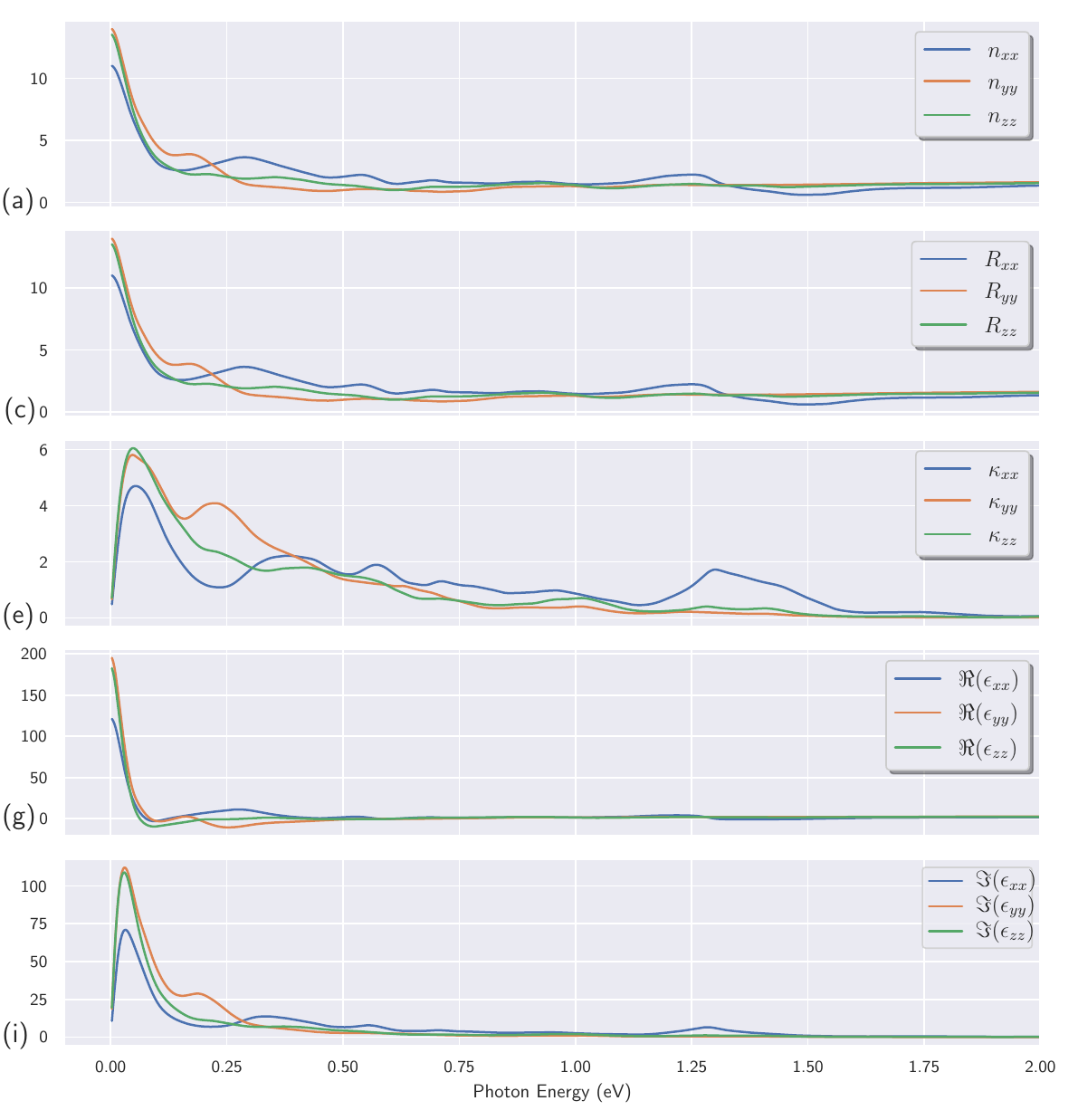}}\hfill
\subfloat{\includegraphics[width=0.5\textwidth]{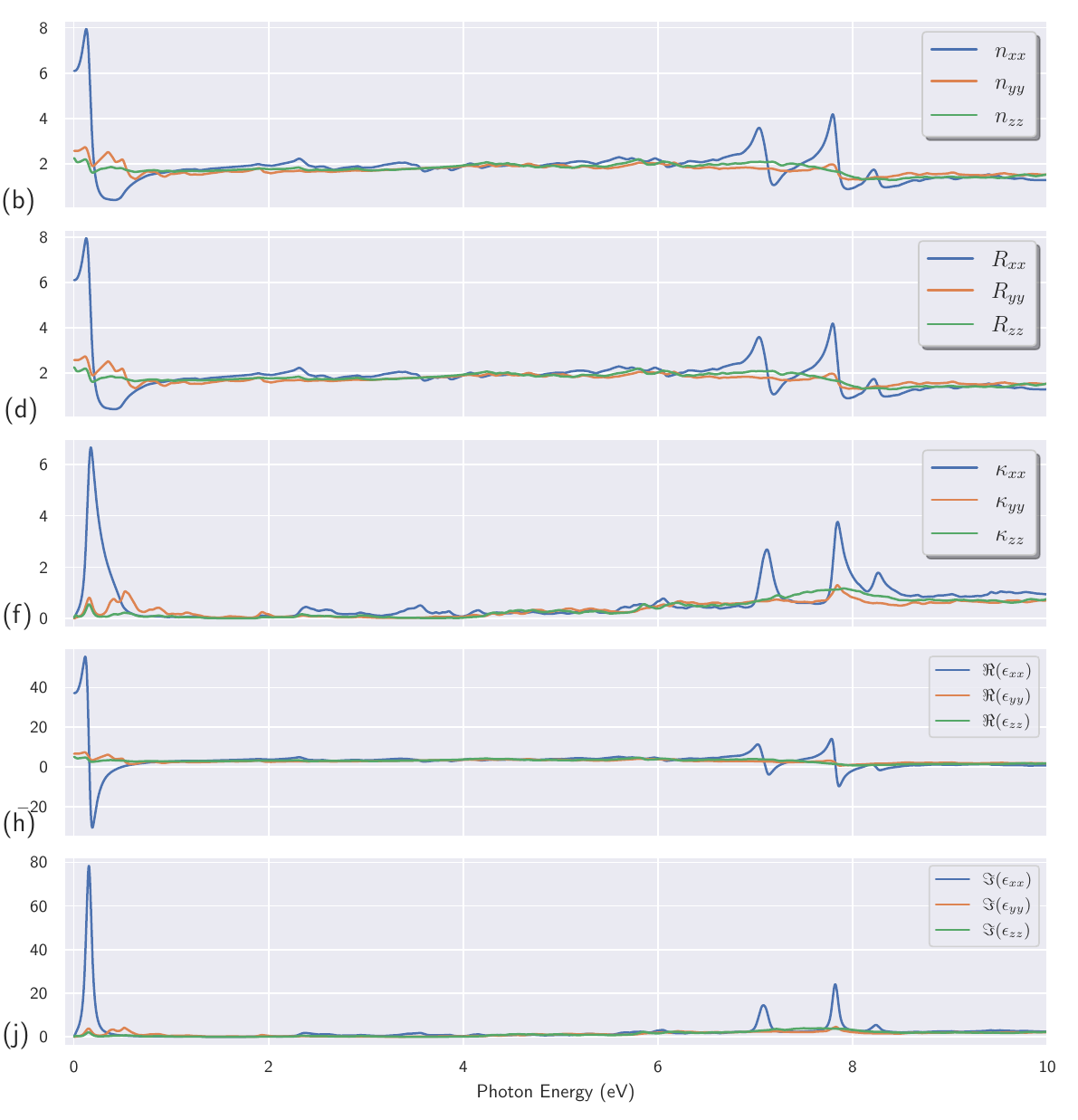}}
\caption{Optical properties of \mol. The left panel displays the optical properties of \PhV, while the
right panel presents the optical properties of \PhI. Subplots (a) and (b) illustrate the reflective indexes, (c) and (d) depict the
refractivities, (e) and (f) demonstrate the extinction coefficient k, (g) and (h) present the real part of the dielectric function, and
(i) and (j) illustrate the imaginary part of the dielectric function.}\label{fig-Optiv}
\end{figure}

It is well established that the imaginary part of the dielectric function $\varepsilon_2(\omega)$ is closely
related to the electronic band structure and exhibits light absorption in the material \cite{RIZWAN2021a}.
Conversely, the real part of the dielectric function $\varepsilon_1(\omega)$ refers to the dispersion of the incident
photons by crystals \cite{ABSIKE2019} and provides information regarding polarization \cite{RIZWAN2021b}. The calculated
static dielectric constants $\varepsilon_1(\omega)$ are equal to \num{121.0460}, and \num{55.5359} for \PhV and \PhI
of the compound, respectively (see Figures \ref{fig-Optiv} g, h). As shown by the results, $\varepsilon_1(\omega)$ reaches a
maximum at $\qty{0.0027}{\electronvolt}$ for \PhV. For \PhI, $\varepsilon_1(\omega)$ reaches a maximum at $\qty{0.1170}{\electronvolt}$, then
decreases and becomes negative reaching a minimum at $\qty{0.1905}{\electronvolt}$. Figures \ref{fig-Optiv} i, j show a good
correspondence between the imaginary part $\varepsilon_2(\omega)$ and the absorption peaks. For \PhI, the
dielectric function increases sharply after $\qty{7}{\electronvolt}$ and $\qty{8}{\electronvolt}$, giving the absorption edge. This
threshold value comes mainly from the electronic transitions between the maxima of the
valence band and the minima of the forbidden conduction band. Usually, it refers to the energy
of the forbidden band, which agrees with the total density of this Phase. This result indicates
that there is a reduction in the band gap, which is in agreement with the results found in
electronic properties (\Cref{tab-ener}).

Among the essential properties for photovoltaic applications is the reflectivity and the
capacity of a sample surface to reflect incident radiation flux. Figures \ref{fig-Optiv} c, d illustrate that
reflectivity $R(\omega)$ decreases with an increase in wavelength, with the highest value of reflectivity
which is \num{12} for \PhV and \num{8} for \PhI.

The extinction coefficient for the $xx$ direction is described in Figures \ref{fig-Optiv} e, f. They
demonstrate that the extinction coefficient is high at low photon energies and decreases as the
energy increases. The values of the peaks for \PhV are \num{4.7013} , \num{2.2148}, \num{1.8937} and \num{1.7215}
at an energy range of \num{0} to $\qty{1.7}{\electronvolt}$, indicating specific absorption bands. While the peaks for
\PhI are \num{6.6514}, \num{2.6889}, \num{3.7611} and \num{1.7846} at an energy range of \num{0} to $\qty{8.5}{\electronvolt}$. A high
extinction coefficient at low photon energies indicates significant absorption, which
corresponds to electronic transitions or interband absorption. Additionally, these peaks
represent resonant absorption where the material absorbs light more effectively at specific
energies.

These results indicate that \mol exhibits strong interaction with
light at low photon energies, with high refractive indices, high reflectivity, and significant
absorption. As the photon energy increases, these optical properties indicate reduced
interaction, with lower refractive indices, lower reflectivity, and reduced absorption. These
properties are consistent with typical dielectric materials, showing strong responses at lower
energies and stabilizing at higher energies, making them potentially useful in applications
where strong light interaction at low energies is desired, such as in optical coatings, sensors, or
devices that require high refractive indices and strong absorption. The decrease in these
properties at higher energies suggests good transparency and reduced reflection, which is
beneficial for optoelectronic applications.

\subsection{Thermoelectric properties}\label{subsec-Ther}

We have employed BoltzTrap2 software \cite{BoltzTraP2} to obtain thermo-electric properties of the
two phases (\PhV and \PhI) of \mol with a constant relaxation
time $\tau_0$. In order to ascertain the thermoelectric (TE) behaviour of both \PhV and \PhI in
detail, the following characteristics have been expressed as a function of chemical potential:
the Seebeck coefficient ($S$), electrical conductivity ($\sigma$), power factor ($PF$), electronic specific
heat ($Cv$), electronic thermal conductivity ($\kappa$), number of concentration carriers ($n$) and density
of states (DOS). The calculated characteristics have been plotted for a range of temperatures
between \num{200} and $\qty{600}{\kelvin}$.

The thermoelectric properties of any material are determined by its figure of merit, which
is expressed as \Cref{eq-therm}:
\begin{equation}
 ZT=\frac{1}{\kappa}(S^2\sigma T),
 \label{eq-therm}
\end{equation}
where:
\begin{itemize}
 \item  the Seebeck coefficient, which is a measure of a material's ability to generate a voltage
difference from a temperature difference, is represented by the symbol $S$,
\item  the symbol $\sigma$ represents electrical conductivity,
\item the symbol $\kappa$ represents the thermal conductivity,
\item the symbol $T$ represents the absolute temperature.
\end{itemize}

\begin{figure}[htbp]
 \centering
 \leavevmode 
 \subfloat[Seebeck coefficient ($S$)]{\includegraphics[width=0.5\textwidth]{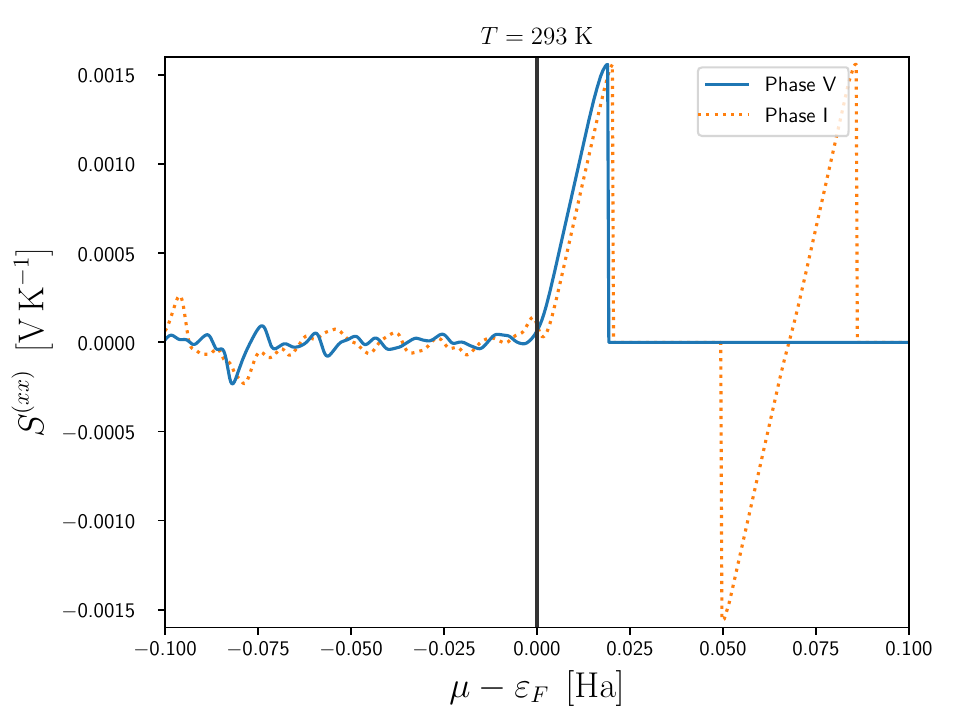}}\hfill
 \subfloat[Electric conductivity rate ($\sigma/\tau_0$)]{\includegraphics[width=0.5\textwidth]{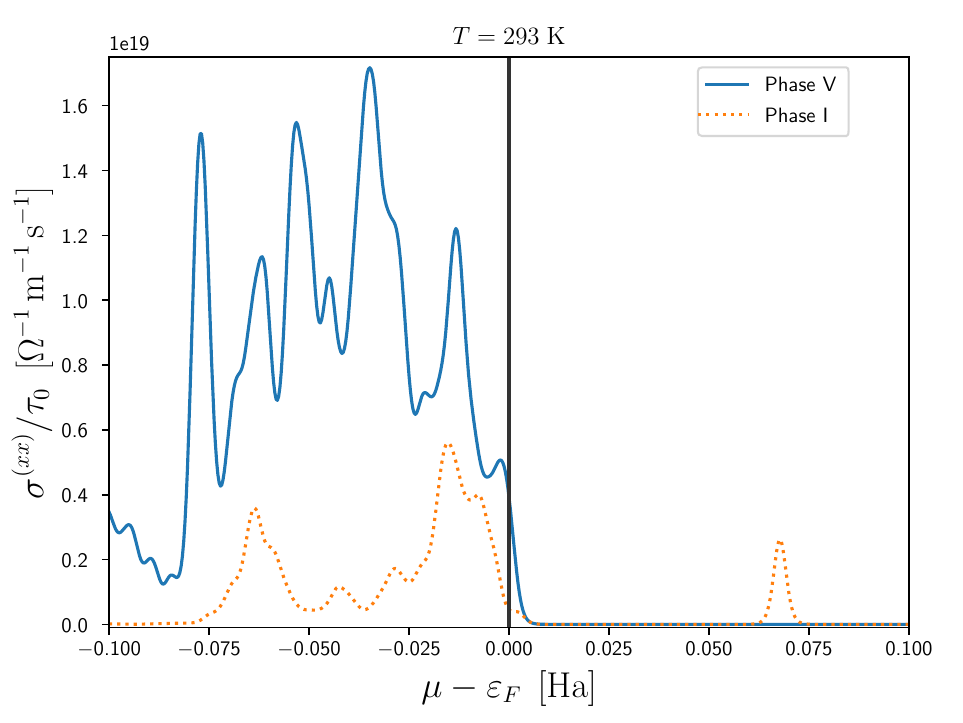}}\\
 \subfloat[Electronic thermal conductivity ($\kappa_e/\tau_0$)]{\includegraphics[width=0.5\textwidth]{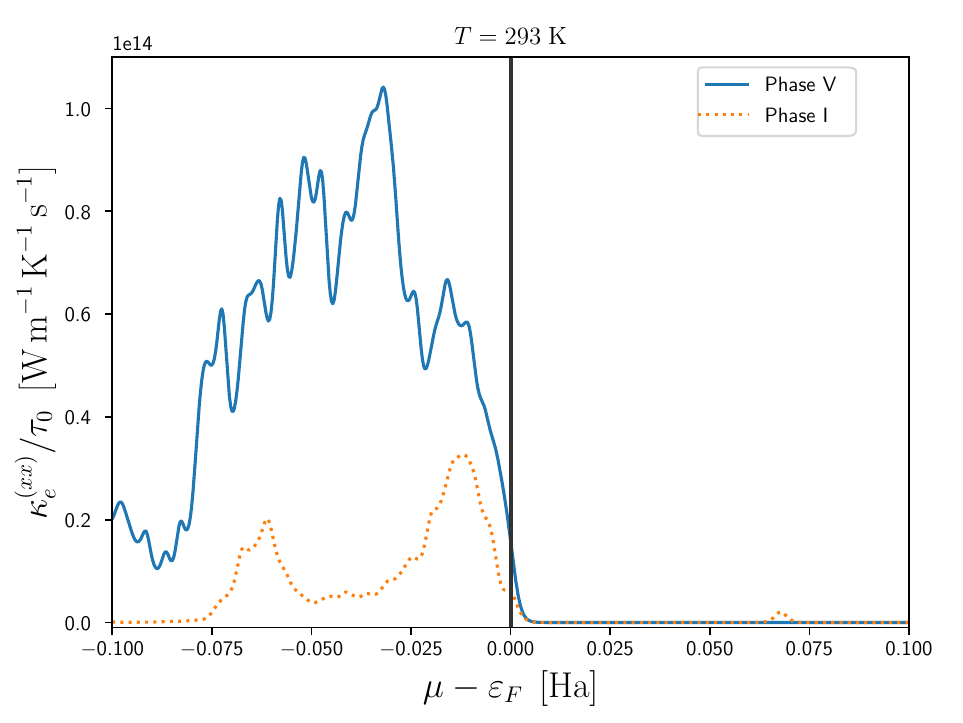}}\hfill
 \subfloat[Power factor ($PF$)]{\includegraphics[width=0.5\textwidth]{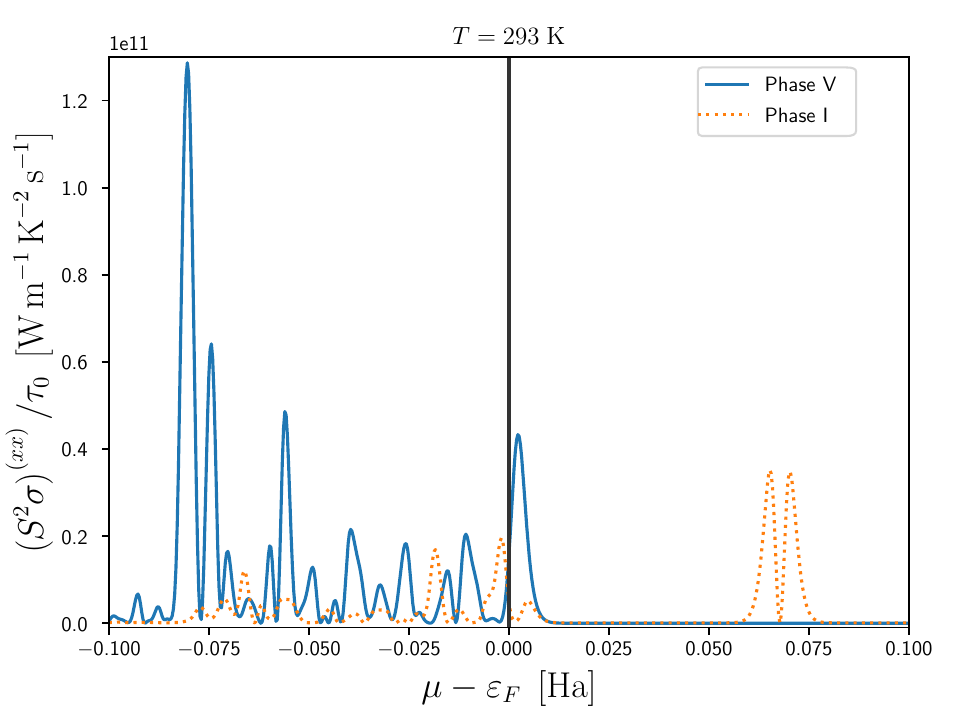}}
 \caption{Transition coefficients as a function of different ground state chemical potential at room temperature for the
\PhV (blue line) and \PhI (dotted orange line) compounds.}
\label{fig-TransC}
\end{figure}

As overall thermal conductivity increases, the figure of merit decreases, and $S$ increases.
A $ZT$ value of at least one is often considered a benchmark for thermoelectric
materials because it signifies that the material has favorable electrical and thermal properties
for efficient energy conversion. Achieving a $ZT$ greater than one is even more desirable,
indicating higher efficiency. Furthermore, a material with a high $ZT$ value is capable of
achieving the optimal energy transformation efficiency, which is typically above $\qty{25}{\percent}$ at the
optimal operating temperature \cite{ALRAHAMNEH2019}.

In Figure \ref{fig-TransC} b, the electric conductivity as a function of chemical potential at room
temperature first increases for both phases. It reaches its global maximum value. Then it
decreases and reaches its global minimum after the null value of the chemical potential at $\mu-\varepsilon_F
= \qty{0.0193}{Ha}$ for \PhV and $\mu-\varepsilon_F = \qty{0.0206}{Ha}$ for \PhI.

The Figure \ref{fig-TransC} c displays the electronic thermal conductivity as a function of chemical
potential. The behavior of the electronic thermal conductivity curves is nonlinear increasing for
both compounds. The maximal electronic thermal conductivity of \PhV at RT is equal to $\qty{1.0416e14}{\watt\per\meter\per\kelvin}$, while it is 
$\qty{0.3264e14}{\watt\per\meter\per\kelvin}$ for \PhI.

The Figure \ref{fig-TransC} d shows how the power factor ($PF$) curves changed when the chemical
potential increased for both compounds. They reach their maximum values of $\qty{1.2868e11}{\watt\per\meter\per\kelvin\squared\per\second}$ for \PhV and 
$\qty{0.3249e11}{\watt\per\meter\per\kelvin\squared\per\second}$ for \PhI. The findings indicate that the
compound in \PhV exhibits superior performance compared to the compound in \PhI,
contingent upon the utilization of an appropriate element dopant at specified concentrations
and/or through band engineering or transition to nanostructures. It is feasible to synthesize
perovskite at the nanoscale, as evidenced by the reduction in lattice thermal conductivity. This
will enhance the augmentation of thermoelectric efficiency.

\begin{figure}[htbp]
 \centering
 \leavevmode
 \subfloat[\PhV]{\includegraphics[width=0.5\textwidth]{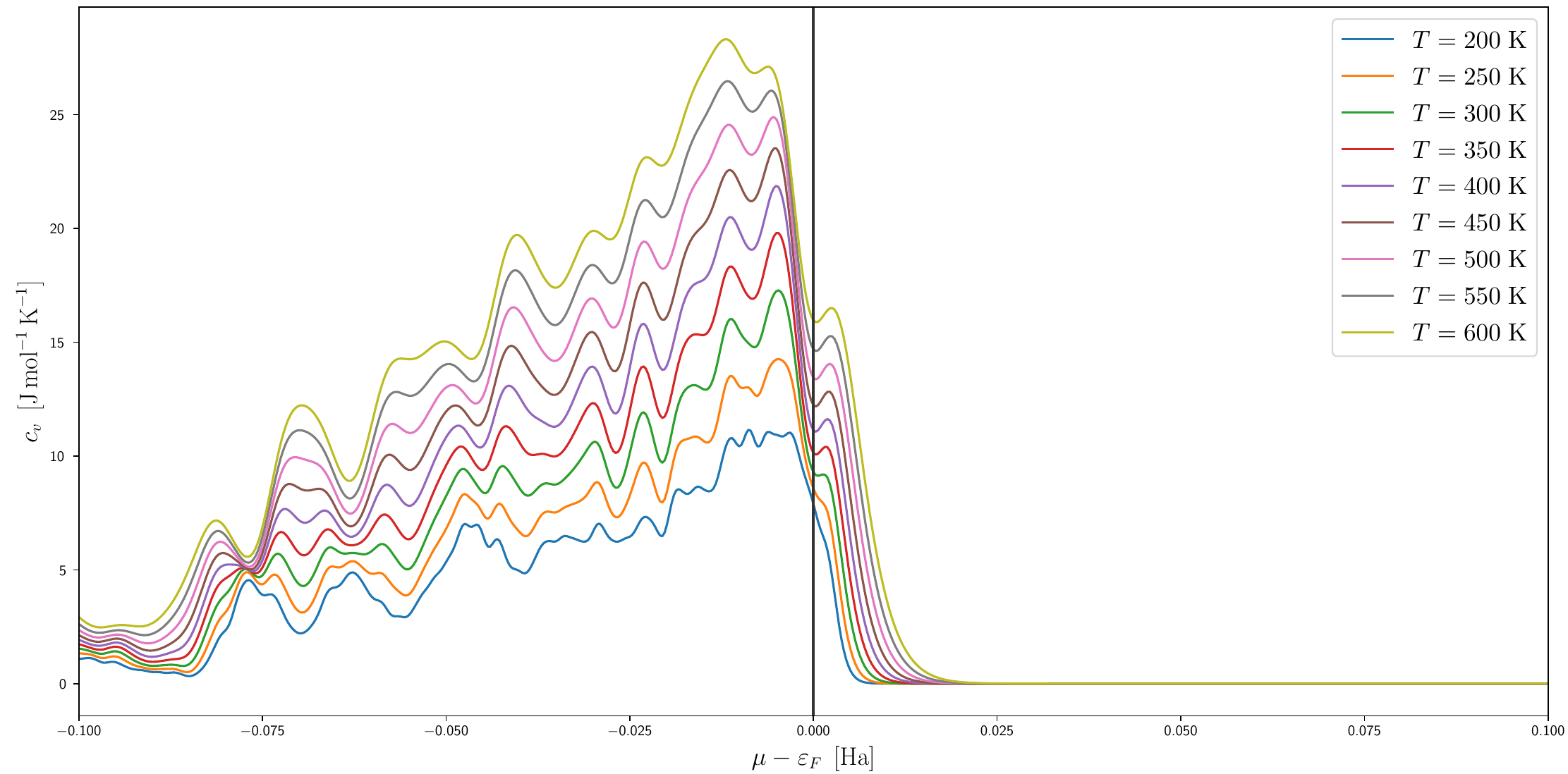}}\hfill
 \subfloat[\PhI]{\includegraphics[width=0.5\textwidth]{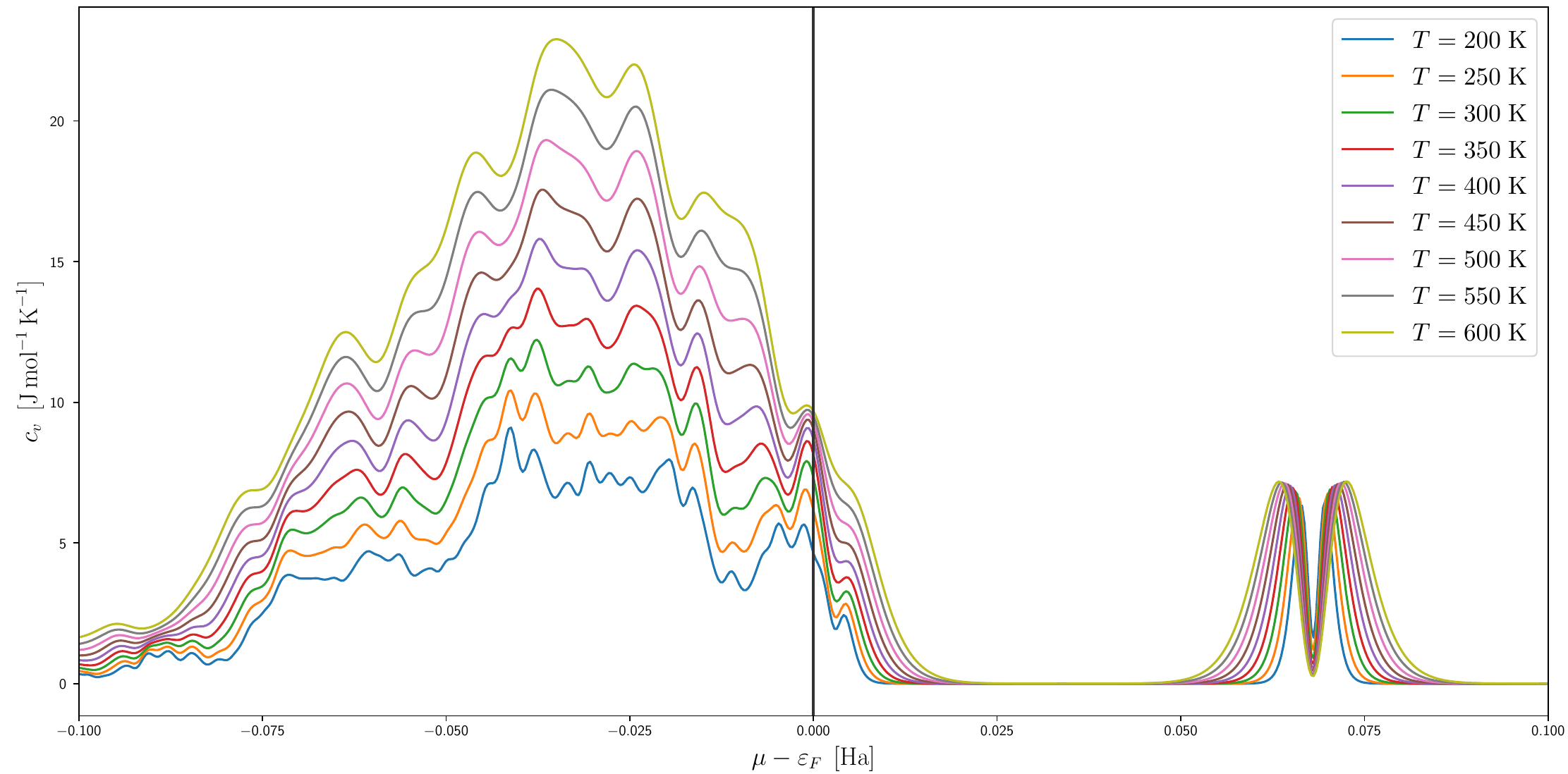}}
 \caption{The electronic specific heat as a function of chemical potential at different temperatures.}\label{fig-cvT}
\end{figure}

\Cref{fig-cvT} illustrates the electronic-specific heat capacity as a function of chemical potential
at different temperatures. The calculated maximum value of the specific heat capacity of \PhV in this study is $\qty{16.8842}{\joule\per\mole\per\kelvin}$ and 
$\qty{11.9643}{\joule\per\mole\per\kelvin}$ for \PhI at RT. The $Cv$ values
for \PhV are slightly greater than those of \PhI.

\begin{figure}[htbp]
 \centering
 \leavevmode
 \subfloat[\PhV]{\includegraphics[width=0.5\textwidth]{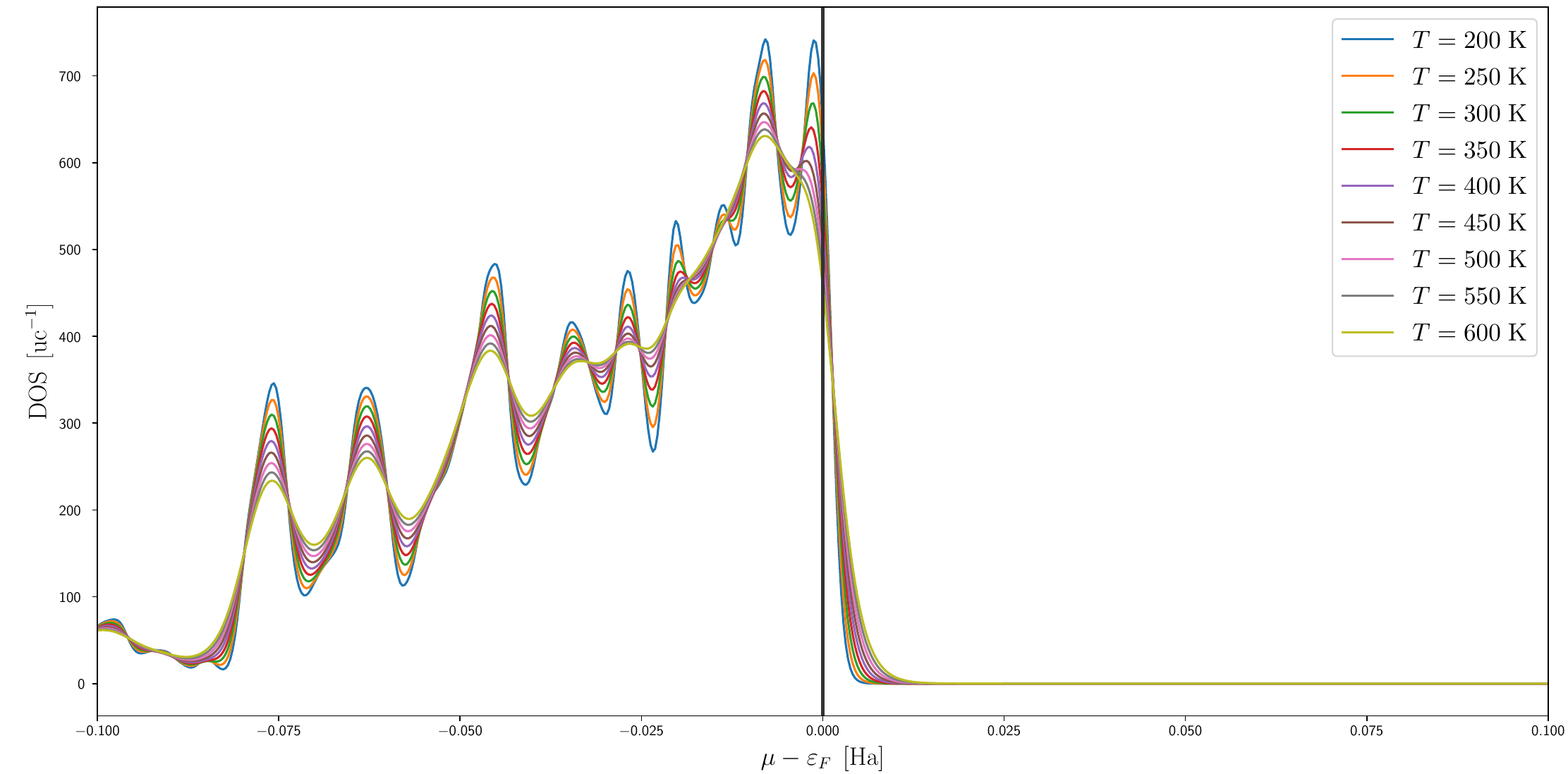}}\hfill
 \subfloat[\PhI]{\includegraphics[width=0.5\textwidth]{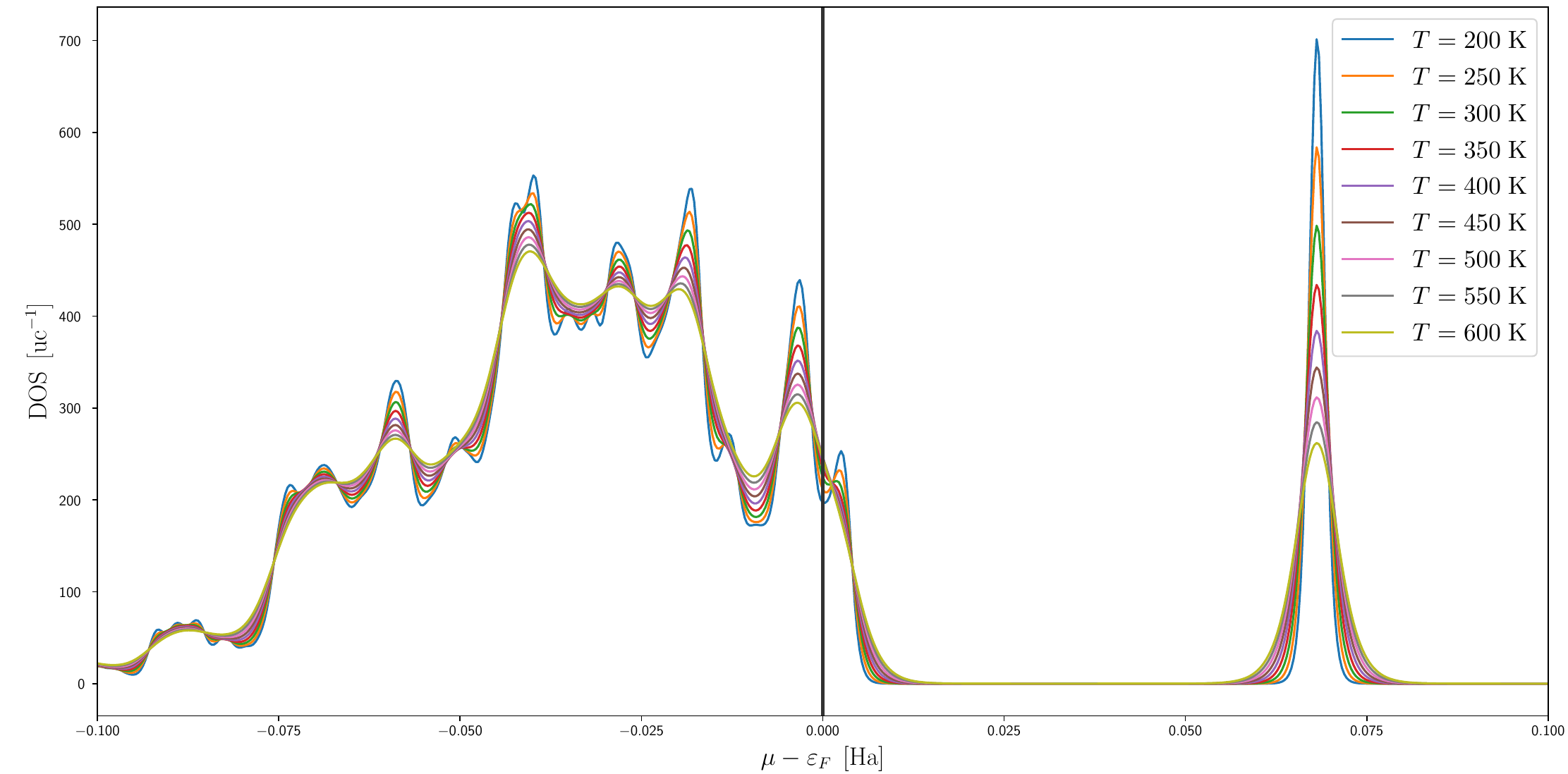}}
 \caption{The density of state at Fermi level DOS ($uc^{-1}$) as a function of chemical potential at different temperatures.}\label{fig-DOST}
\end{figure}

\Cref{fig-DOST} illustrates the variation in the density of the state around the Fermi level as a
function of temperature. For the two alloys, \PhV and \PhI, the band gap persists over a
wide range of temperatures.

\begin{figure}[htbp]
 \centering
 \leavevmode
 \subfloat[\PhV]{\includegraphics[width=0.5\textwidth]{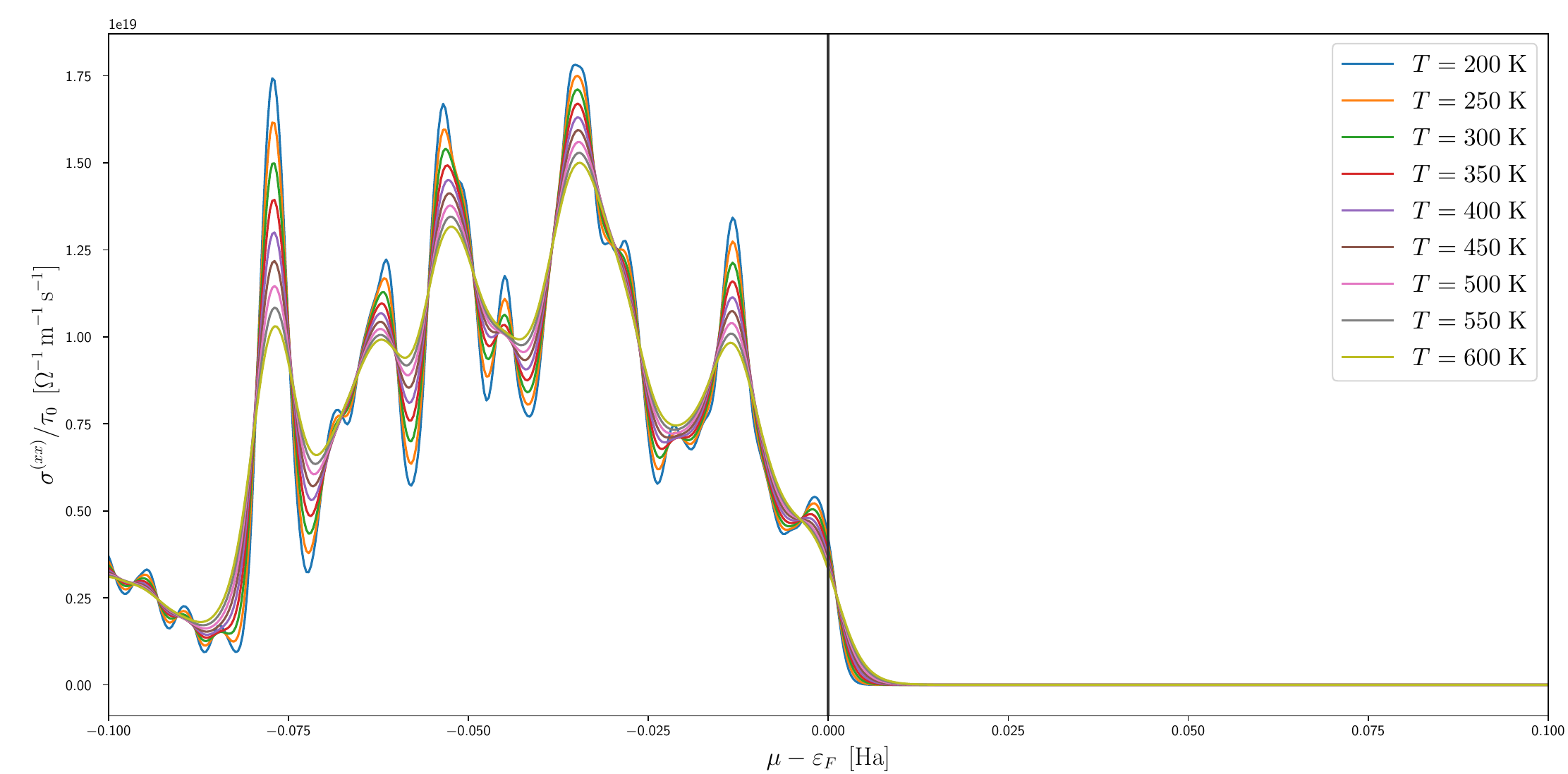}}\hfill
 \subfloat[\PhI]{\includegraphics[width=0.5\textwidth]{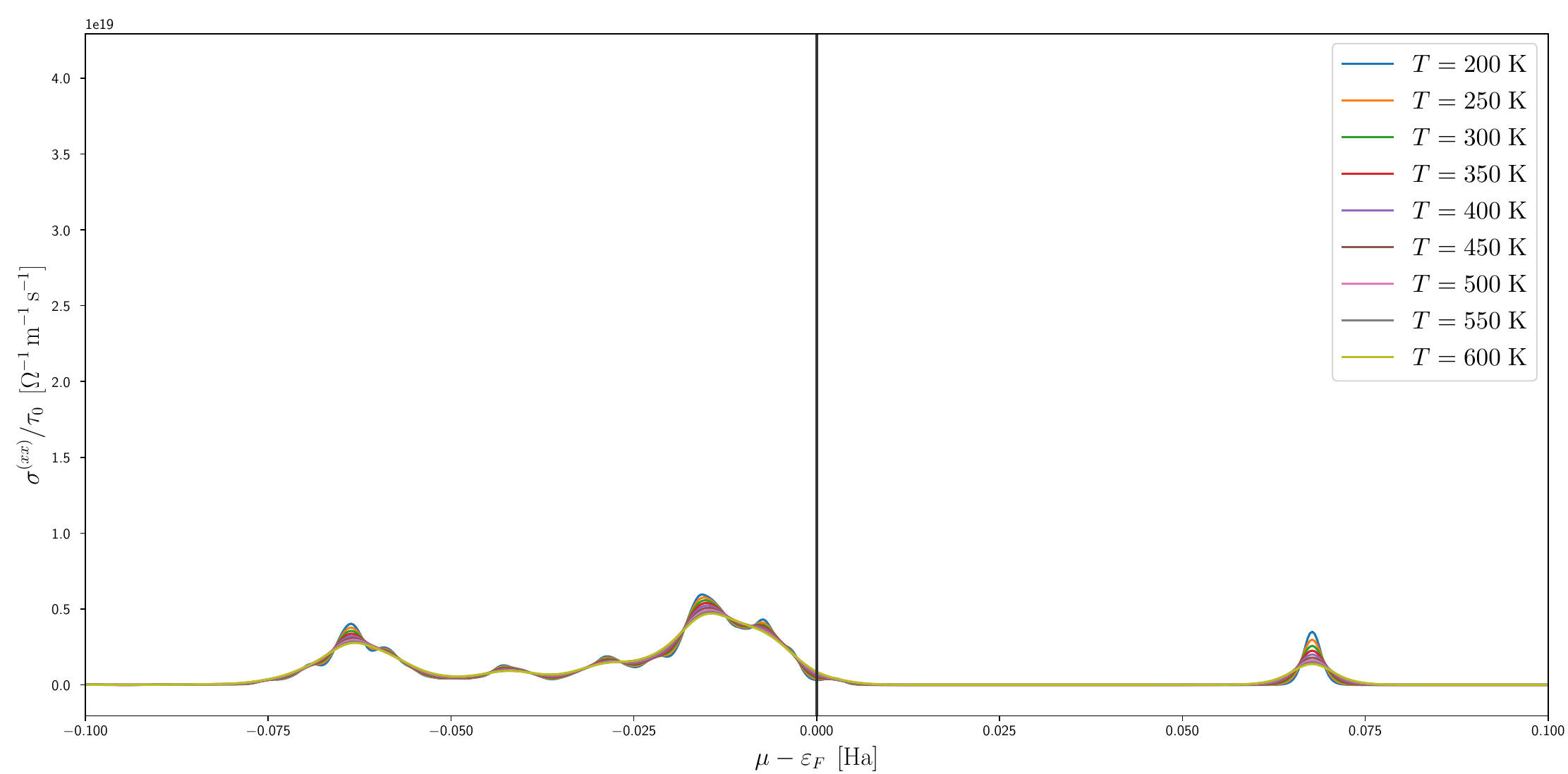}}
 \caption{Electrical conductivity rate ($\sigma/\tau$) as a function of chemical potential at different temperatures.}\label{fig-Sigm}
\end{figure}

\Cref{fig-Sigm} presents the electrical conductivity in \PhV (left panel) and \PhI (right
panel). In \PhV, the electrical conductivity of \mol exhibits multiple
peaks at different temperatures. These peaks suggest changes in conductivity associated with
Phase transitions or structural modifications. The complex behavior likely arises from
interactions between the \ionN anions and the organic cations \ionP. In \PhI,
there is a single prominent peak in the electrical conductivity which means that the overall
conductivity is significantly lower compared to \PhV. This finding indicates that \PhI has different conductive properties, due to a distinct arrangement of 
ions and altered crystal
structure caused by the effect of temperature. Structural Considerations like the structural
dynamics of \mol likely play a crucial role in determining its electrical
behavior. And Changes in hydrogen bonding (such as \ce{N-H-Cl} interactions) and the
arrangement of \ionN anions contribute to the observed conductivity variations.

\begin{figure}[htbp]
 \centering
 \leavevmode
 \subfloat[\PhV]{\includegraphics[width=0.5\textwidth]{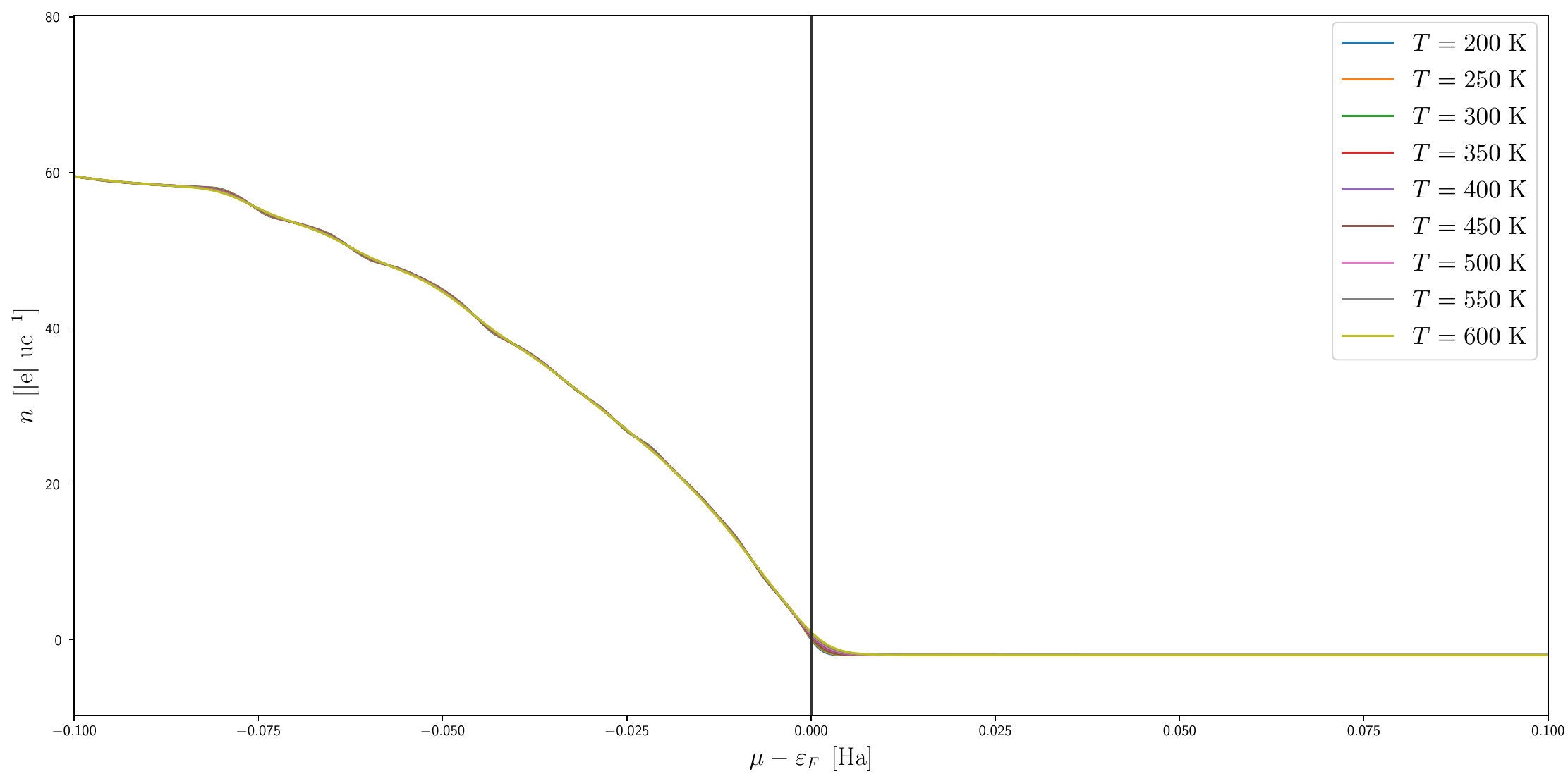}}\hfill
 \subfloat[\PhI]{\includegraphics[width=0.5\textwidth]{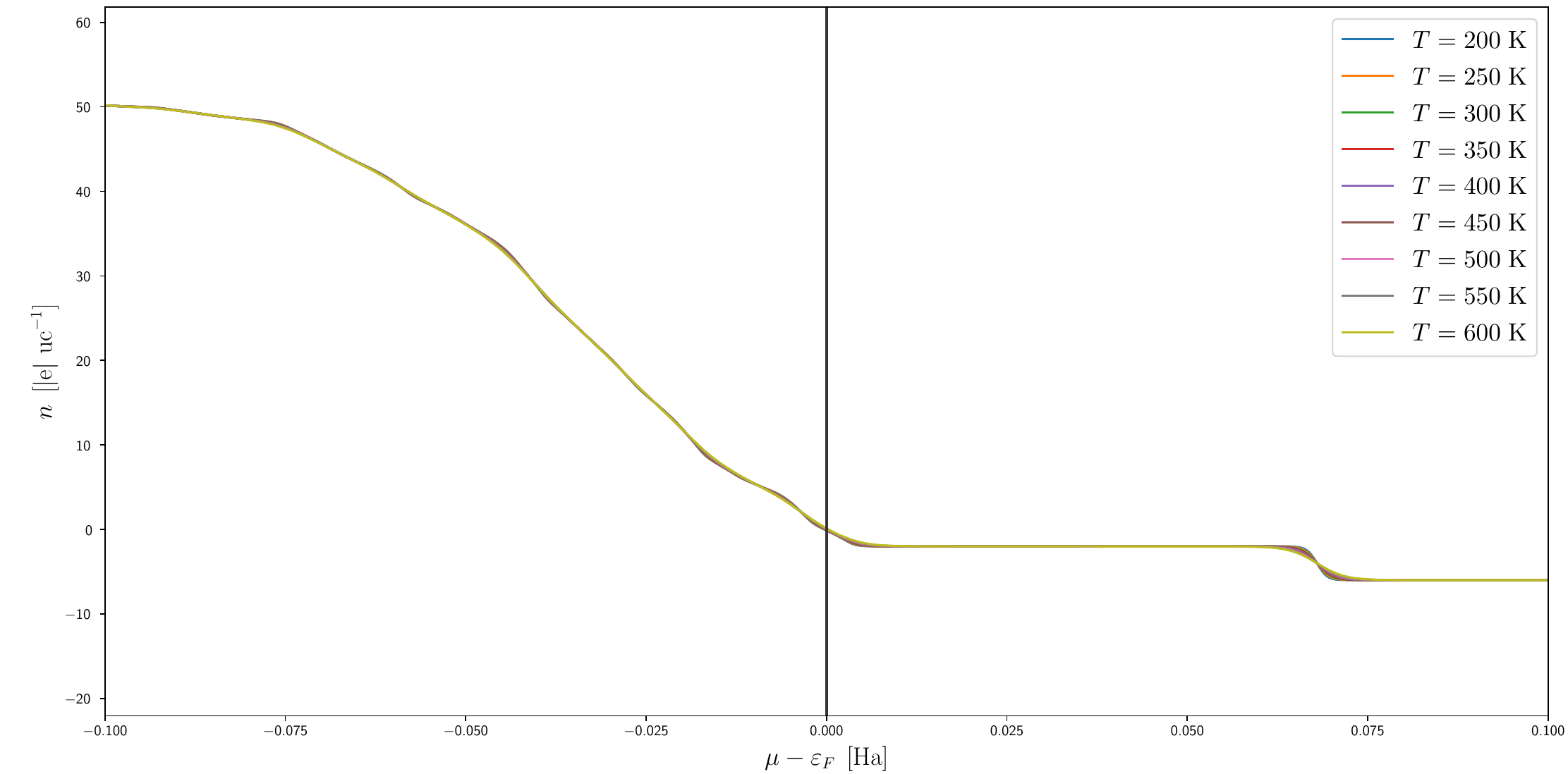}}
 \caption{Number of carrier concentration as a function of chemical potential at different temperatures. The efficiency
of a thermoelectric device ($n$) is not determined by maximum $ZT$ at a single temperature but depends on the average $ZT$
over a wide temperature range following.}\label{fig-Carrier}
\end{figure}

\Cref{fig-Carrier} illustrates the carrier concentration in the two phases. The carrier concentration
for both phases appears to be relatively constant across the distance measured. The carrier
concentration for \PhI is slightly higher than that of \PhV. As is well known, the carrier
concentration is typically used to refer to the number of charge carriers per unit volume in a
semiconductor. In this context, a charge carrier could be an electron or a hole. The results
demonstrate that the temperature at which the measurement is taken affects the carrier
concentration.

\section{Conclusion}

This inquiry, utilizing Density Functional Theory (DFT) computations, has thoroughly
investigated the structural, electronic, optical, and thermoelectric attributes of the hybrid
organic-inorganic perovskite \mol in its \PhV and \PhI
configurations. Our results emphasize the considerable impact of phase transitions on the
inherent characteristics of this material.

Structural optimization employing GGA-PBE produced lattice parameters in remarkable
concordance with experimental observations, corroborating the validity of our computational
methodology. A notable decrease in the band gap energy from $\qty{4.020}{\electronvolt}$ in \PhV to $\qty{1.710}{\electronvolt}$
in \PhI was detected, signifying enhanced electronic conductivity. Optical properties
assessment disclosed augmented light absorption and interaction, particularly in \PhI,
implying its potential for photovoltaic applications. Evaluation of thermoelectric properties
illustrated the superior performance of \PhV, rendering it a promising candidate for
thermoelectric devices. The distinct electrical conductivity profiles of the two phases highlight
the effect of structural dynamics and temperature on the material's conductive behavior.

In summary, the observed variability in electronic and optical properties across the phases of
\mol establishes this compound as a multifaceted material for
optoelectronic and thermoelectric applications. Future inquiries could concentrate on
investigating doping strategies and nanoscale synthesis to further enhance its properties.

\backmatter

\bmhead{Acknowledgements}

This research was supported through computational resources of HPC-MARWAN
(hpc.marwan.ma) provided by the National Center for Scientific and Technical Research (CNRST), Rabat, Morocco.

\section*{Declarations}

\begin{itemize}
\item Conflict of interest/Competing interests:
On behalf of all authors, the corresponding author states that there is no conflict of interest. 
\item Data availability:
Data sets generated during the current study are available from the corresponding author on reasonable request.
\item Code availability:
Code used during the current study are available from the corresponding author on reasonable request. 
\item Author contribution:
All authors contributed to the study conception and design. Code preparation, data collection and analysis were performed by Hafida ZIOUANI and Jean-Pierre 
TCHAPET NJAFA. The first draft of the manuscript was written by Hafida ZIOUANI and all authors commented on previous versions of the 
manuscript. All authors read and approved the final manuscript.
\end{itemize}

\bigskip

\bibliography{Biblio}

\end{document}